\documentclass[aps,prb,twocolumn,showpacs,superscriptaddress]{revtex4-1}
\usepackage[colorlinks=true,linkcolor=Maroon,citecolor=OliveGreen,urlcolor=Blue,linktoc=page]{hyperref}
\usepackage[usenames,dvipsnames]{xcolor}

\usepackage{latexsym}
\usepackage{ifpdf}
\usepackage{graphicx}
\usepackage{epsfig}
\usepackage{amsmath, amssymb, bbm, mathrsfs}
\usepackage{multirow}
\usepackage[varg]{txfonts}
\usepackage{ulem}
\usepackage{fancyhdr}
\usepackage{color}
\usepackage{bm}
\usepackage{epsfig}
\usepackage{eurosym}
\usepackage{subfigure}
\usepackage{comment}
\usepackage[colorlinks=true]{hyperref}
\usepackage{scrhack}
\usepackage{physics}

\renewcommand{\Re}{\mbox{Re} }
\renewcommand{\Im}{\mbox{Im} }

\def\S{\textbf{\textit{S}}}

\def\H{\mathcal{H}}
\def\PT{\mathcal{PT}}

\def\C{\mathbb{C}}

\def\Heff{\textbf{\textit{H}}_\text{eff}}
\def\HH{\textbf{\textit{H}}}

\def\ep{\textbf{\textit{e}}_\textbf{\textit{p}}}

\def\j{\textbf{\textit{j}}}
\def\muB{\mu_\text{B}}

\begin{document}
	\title{Parity-time symmetry breaking in magnetic systems}
	\author{Alexey Galda}
	\author{Valerii M. Vinokur}
	\affiliation{Materials Science Division, Argonne National Laboratory,
		9700 South Cass Avenue, Argonne, Illinois 60439, USA}
	\date{\today}
\begin{abstract}
	The understanding of out-of-equilibrium physics, especially dynamic instabilities and dynamic phase transitions, is one of the major challenges of contemporary science, spanning the broadest wealth of research areas that range from quantum optics to living organisms. Focusing on nonequilibrium dynamics of an open dissipative spin system, we introduce a non-Hermitian Hamiltonian approach, in which non-Hermiticity reflects dissipation and deviation from equilibrium. The imaginary part of the proposed spin Hamiltonian describes the effects of Gilbert damping and applied Slonczewski spin-transfer torque. In the classical limit, our approach reproduces Landau-Lifshitz-Gilbert-Slonczewski dynamics of a large macrospin. We reveal the spin-transfer torque-driven \textit{parity-time} symmetry-breaking phase transition corresponding to a transition from precessional to exponentially damped spin dynamics. Micromagnetic simulations for nanoscale ferromagnetic disks demonstrate the predicted effect. Our findings can pave the way to a general quantitative description of out-of-equilibrium phase transitions driven by spontaneous parity-time symmetry breaking.
\end{abstract}
	
\maketitle

\section*{Introduction}
A seminal idea of parity-time ($\PT$)-symmetric quantum mechanics~[\onlinecite{Bender98},\onlinecite{Bender99}], that has stated that the condition of Hermiticity in standard quantum mechanics required for physical observables and energy spectrum to be real can be replaced by a less restrictive requirement of invariance under combined parity and time-reversal symmetry, triggered an explosive development of a new branch of science. The interpretation of $\PT$ symmetry as ``balanced loss and gain"~[\onlinecite{Ruschhaupt}] connected $\PT$ symmetry breaking to transitions between stationary and nonstationary dynamics and established its importance to understanding of the applied field-driven instabilities. Experiments on a diverse variety of strongly correlated systems and phenomena including optics and photonics~[\onlinecite{Klaiman, Longhi09, Longhi10, Schomerus, Ramezani, Rueter, Sheng}], superconductivity~[\onlinecite{Rubinstein07, Rubinstein10, Vinokur12}], Bose-Einstein condensates~[\onlinecite{Graefe08}],	nuclear magnetic resonance quantum systems~[\onlinecite{Zheng}], and coupled electronic and mechanical oscillators~[\onlinecite{Bender13a, Bender13b, Schindler}] revealed $\PT$ symmetry-breaking transitions driven by applied fields. These observations stimulated theoretical focus on far-from-equilibrium instabilities of many-body systems~[\onlinecite{Rubinstein10},\onlinecite{Vinokur12},\onlinecite{Science}] that are yet not thoroughly understood.

Here we demonstrate that the non-Hermitian extension of classical Hamiltonian formalism provides quantitative description of dissipative dynamics and dynamic phase transitions in out-of-equilibrium systems. Focusing on the case of spin systems, we consider the zero-temperature spin dynamics under the action of basic nonconservative forces: phenomenological Gilbert damping~[\onlinecite{Gilbert}] and Slonczewski spin-transfer torque~[\onlinecite{Slon96}] (STT). The latter serves as the most versatile way of directly manipulating magnetic textures by external currents. We propose a general complex spin Hamiltonian, in which Slonczewski STT emerges from an imaginary magnetic field. The $\PT$-symmetric version of the Hamiltonian is shown to exhibit a phase transition associated with inability of the system to sustain the balance between loss and gain above a certain threshold of external nonconservative field.

In the classical limit of a large spin, our formalism reproduces the standard Landau-Lifshitz-Gilbert-Slonczewski~[\onlinecite{LL, Gilbert, Slon96}] (LLGS) equation of spin dynamics and predicts the $\PT$ symmetry-breaking phase transition between stationary (conservative) and dissipative (nonconservative) spin dynamics. In this Letter we focus on a single spin, yet our theory can be extended to coupled spin systems in higher dimensions. Moreover, as spin physics maps onto a wealth of strongly correlated systems and phenomena ranging from superconductivity to cold-atom and two-level systems, our results provide quantitative perspectives on the nature of phase transitions associated with $\PT$ symmetry breaking in a broad class of far-from-equilibrium systems.

\begin{figure}[!t]
	\includegraphics[width=1\columnwidth]{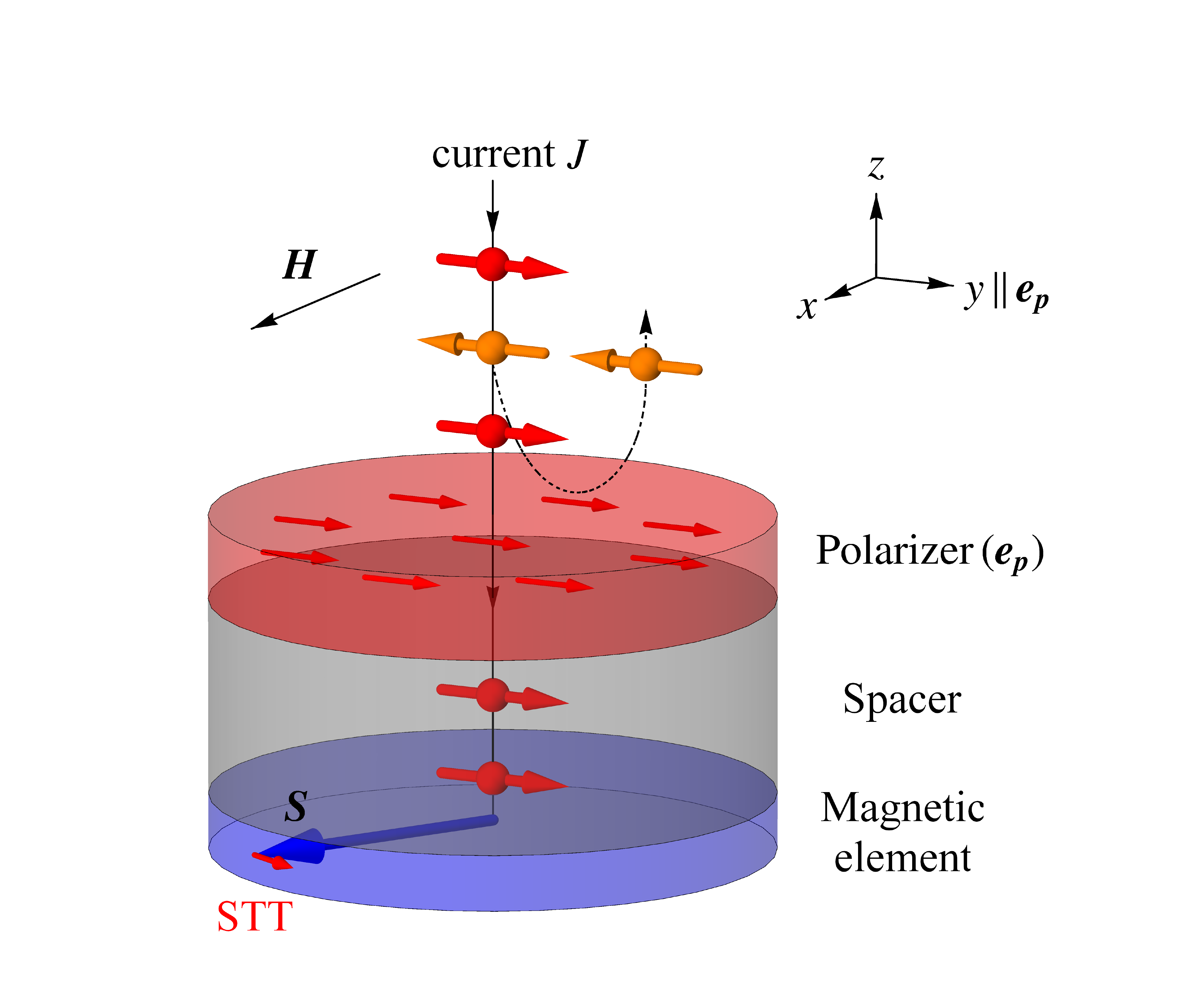}
	\caption{Schematic representation of the system setup. Ferromagnetic cylinder (blue) is placed in magnetic field $\HH$ applied along the $x$ axis, and STT-inducing electric current $J$ is polarized in the direction $\ep$ along the $y$ axis. Spin-polarized current passes through a nonmagnetic metallic spacer and induces torque (Slonczewski STT, shown by the small red arrow) on the total spin $\S$.}
	\label{fig1}
\end{figure}

We introduce the non-Hermitian Hamiltonian for a single spin operator~$\hat\S$:
\begin{equation}\label{H}
	\hat\H = \frac{ E\bigl( \hat\S\bigr) + i\,\j \cdot \hat\S}{1 - i\,\alpha}\,,
\end{equation}
where $E\bigl( \hat\S\bigr)$ denotes the standard Hermitian spin Hamiltonian determined by the applied magnetic field $\HH$ and magnetic anisotropy constants $k_i$ in the ${x,y,z}$ directions: ${E\bigl( \hat\S\bigr) = \sum_{i}k_i\, \hat S_i^2 + \gamma\HH \cdot \hat\S}$. A schematic system setup is shown in Fig.~\ref{fig1}. The phenomenological constant ${\alpha > 0}$ in Eq.~(\ref{H}) describes damping; the imaginary field $i\j$ is responsible for the applied Slonczewski STT, with $\j\,S = \ep\,(\hbar/2e)\eta J$ being the spin-angular momentum deposited per second in the direction $\ep$ with spin polarization ${\eta = (J_\uparrow - J_\downarrow)/(J_\uparrow + J_\downarrow)}$ of the incident current $J$; and ${\gamma = g\muB/\hbar}$ is the absolute value of the gyromagnetic ratio; ${g \simeq 2}$, $\muB$ is the Bohr magneton, $\hbar$ is the Planck's constant, and $e$ is the elementary charge. We conjecture that Eq.~(\ref{H}) serves as a fundamental generalization of the Hamiltonian description of both quantum and classical spin systems, which constitutes one of our core results. This form of the Hamiltonian proves extremely useful for the general understanding of STT-driven dissipative spin dynamics. In this work we focus primarily on the classical limit of spin dynamics, while the semiclassical limit of finite spin will be considered elsewhere.

Spin dynamics in the classical limit is conveniently obtained by studying expectation value of the Hamiltonian~(\ref{H}) with respect to SU(2) spin-coherent states~[\onlinecite{Lieb},\onlinecite{Garg}]: ${\ket{z} = e^{z\,\hat S_+}\!\ket{S, -S}}$, where ${\hat S_\pm \equiv \hat S_x \pm i\hat S_y}$, and ${z \in \C}$ is the standard stereographic projection of the spin direction on a unit sphere, ${z = (s_x + is_y)/(1 - s_z)}$, with ${s_i \equiv S_i/S}$. Note that such parametrization of the phase space for a classical single-spin system (i.e., in the limit $S \to \infty$) guarantees the invariance of the traditional equation of motion~[\onlinecite{Garg}] under generalization to non-Hermitian Hamiltonians (see Appendix A):
\begin{equation}\label{zHamilton}
	\dot z = i\,\frac{ \left(1 + \bar z z\right)^2}{2S} \frac{\partial \H}{\partial \bar z}\,,
\end{equation}
where $z$ and $\bar z$ form a complex conjugate pair of stereographic projection coordinates, and
\begin{equation}
	\H (z, \bar z) = \frac{\expval{\hat\H}{z}}{\bra{z}\ket{z}}
\end{equation}
is the expectation value of the Hamiltonian~$(\ref{H})$ in spin-coherent states (for a detailed review see, e.g., Ref.~[\onlinecite{Radcliffe}]). In this formulation, the eigenstates of $\hat\H$ correspond to the fixed points $z_i$ of the equation of motion for $\H$, while the eigenvalues (i.e., energy values) are equal to $\H$ evaluated at the corresponding fixed points, $E_i = \H(z_i, \bar z_i)$.

Assuming a constant magnitude of the total spin, $\dot S = 0$, Eq.~(\ref{zHamilton}) reduces to the following equation of spin dynamics in the classical limit:
\begin{equation}\label{Re-Im}
	\dot{\S} = \nabla_{\S}\!\left( \Re\, \H\right) \times \S + \frac{1}{S}\left[\nabla_{\S}\!\left( \Im\, \H\right) \times \S \right] \times \S\,.
\end{equation}
Here we refer to the real and imaginary parts of the Hamiltonian function $\H$ written in the spin $\S$ representation. For the non-Hermitian Hamiltonian~(\ref{H}), Eq.~(\ref{Re-Im}) reproduces the LLGS equation describing dissipative STT-driven dynamics of a macrospin:
\begin{align}\label{LLGSs}
	\left(1 + \alpha^2\right)\dot{\S} &= \gamma\Heff \times \S + \frac{\alpha\gamma}{S}[\gamma\Heff \!\times\! \S] \times \S + \frac1{S}\,\S \times [\S \times \j]\notag\\
	&+ \alpha\,\S \times \j\,,\\
	\gamma\Heff &= \nabla_{\S} E(\S)\,.
\end{align}

The first two terms in Eq.~(\ref{LLGSs}) describe the standard Landau-Lifshitz torque and dissipation, while the last two are responsible for the dissipative (antidamping) and conservative (effective field) Slonczewski STT contributions, correspondingly, both of which appear naturally from the imaginary magnetic field term in the Hamiltonian~(\ref{H}).

\section*{$\PT$-symmetric Hamiltonian}
Slonczewski STT turns the total spin-angular momentum, $\S$, in the direction of spin-current polarization, $\ep$, without changing its magnitude. On the $\S$-sphere this can be represented by a vector field converging in the direction of $\ep$ and originating from the antipodal point. It is the \textit{imaginary} magnetic field $i\j$ that produces exactly the same effect on spin dynamics, according to Eq.~(\ref{zHamilton}). The action of STT is invariant under the simultaneous operations of time reversal and reflection with respect to the direction $\ep$, which is the underlying reason behind the inherent $\PT$ symmetry of certain STT-driven magnetic systems, including the one considered below.

\begin{figure*}[!t]
	\begin{subfigure}
		\centering
		\includegraphics[width=\columnwidth]{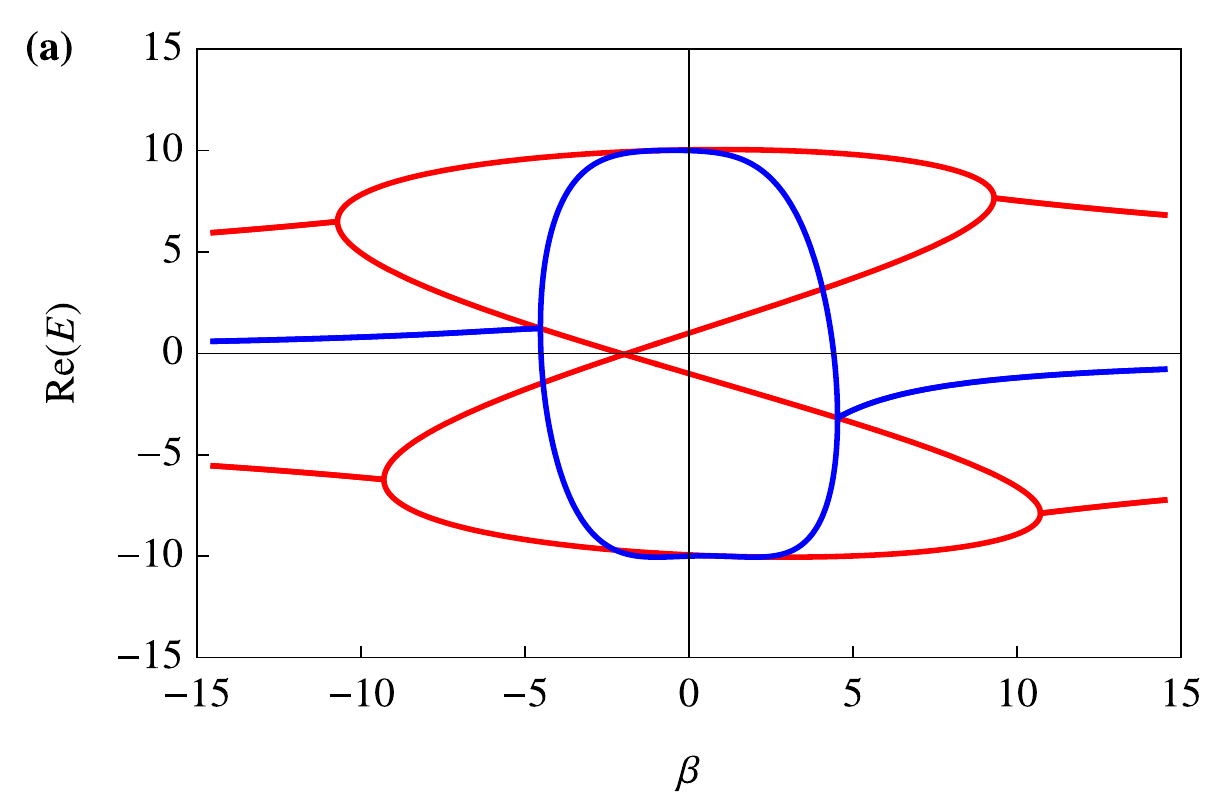}
	\end{subfigure}
	\begin{subfigure}
		\centering
		\includegraphics[width=\columnwidth]{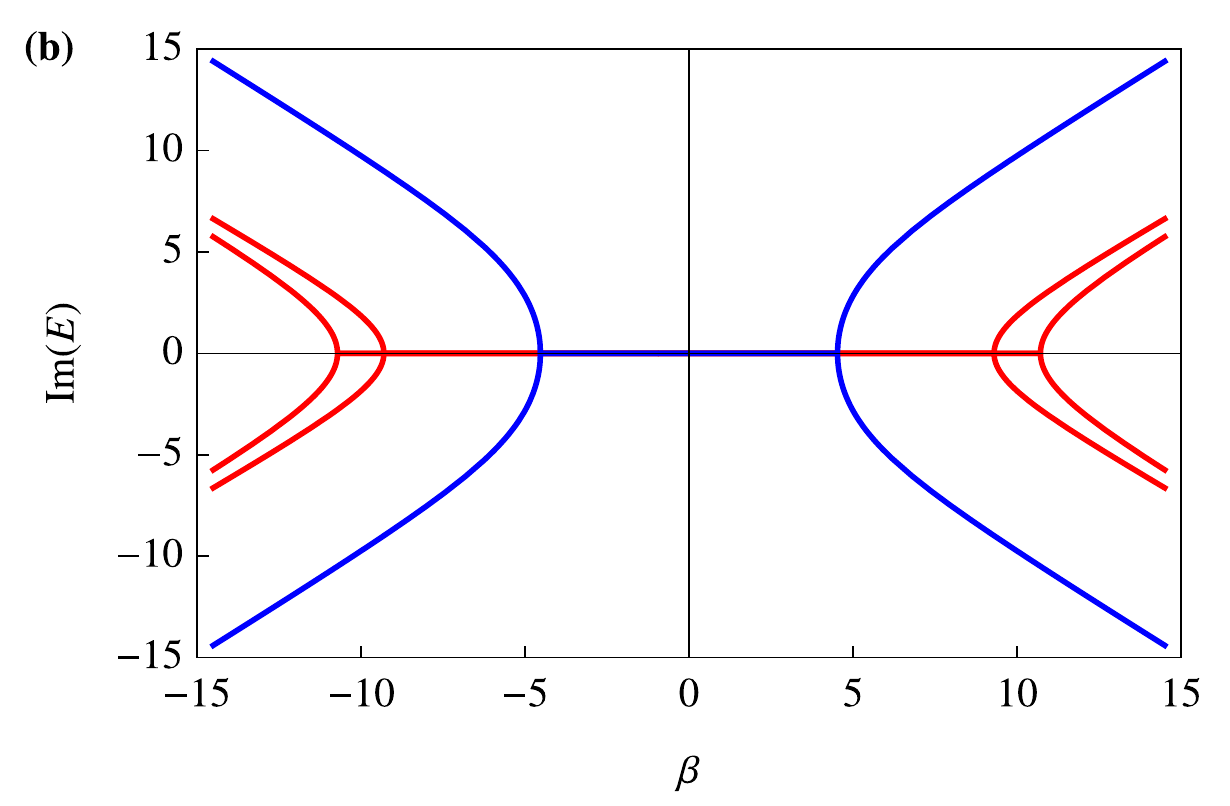}
	\end{subfigure}	
	\caption{Real (a) and imaginary (b) parts of energy spectrum of the Hamiltonian~(\ref{HDz}) as a function of the STT parameter $\beta$ for ${h_x = 1}$ and ${D = 20}$. Blue and red lines correspond to the eigenvalues $E_{1,2}$ and $E_{3-6}$, respectively. The first $\PT$ symmetry-breaking transition occurs at ${|\beta| = \beta_1 \approx 4.5}$.}
	\label{fig2}
\end{figure*}

Before turning to the $\PT$-symmetric form of Hamiltonian~(\ref{H}), we note that $\PT$-symmetric systems play an important role in the studies of nonequilibrium phenomena and provide a unique nonperturbative tool for examining the phase transition between stationary and nonstationary out-of-equilibrium dynamics. We show that despite being non-Hermitian, such systems can exhibit both of the above types of behavior, depending on the magnitude of the external nonconservative force. In the parametric regime of \textit{unbroken} $\PT$ symmetry, systems exhibit physical properties seemingly equivalent~[\onlinecite{Ali}] to those of Hermitian systems: real energy spectrum, existence of integrals of motion (see Appendix C), and, notably, the validity of the quantum Jarzynski equality~[\onlinecite{Deffner}]. However, in the regime of \textit{broken} $\PT$ symmetry, one observes complex energy spectrum and nonconservative dynamics. Therefore, the true transition between stationary and nonstationary dynamics can be identified as the $\PT$ symmetry-breaking phase transition.

Spin systems are generally subject to various nonlinear magnetic fields including ones originating from shape, exchange, and magnetocrystalline and magnetoelastic anisotropies. Restricting ourselves for simplicity to a second-order anisotropy term, we arrive at the following Hamiltonian for a nonlinear magnetic system with uniaxial anisotropy and applied Slonczewski STT:
\begin{equation}\label{HDz}
	\hat \H_\PT =  \gamma H_0 \left(k_z \hat S_z^2 + h_x \hat S_x + i\,\beta\, \hat S_y\right)\,,
\end{equation}
where the applied magnetic field $h_x$ is measured in units of some characteristic magnetic field $H_0$, and $\beta$ is a dimensionless STT parameter determining the relative to $S$ amount of angular momentum transfered in time $\tau \equiv (\gamma H_0)^{-1}$ (characteristic timescale of the dynamics, used as a unit of dimensionless time in what follows). The Hamiltonian~(\ref{HDz}) modeling the dynamics of the free magnetic layer in a typical nanopillar device with fixed polarizer layer (see Fig.~\ref{fig1}) is $\PT$ symmetric: It is invariant under simultaneous action of parity and time-reversal operators (${y \to -y}$, ${t \to -t}$, ${i \to -i}$). Because the Hamiltonian $\hat \H_\PT$ commutes with an antilinear operator $\PT$, its eigenvalues are guaranteed to appear in complex conjugate pairs. Notice that $\PT$-symmetric Hamiltonian~(\ref{HDz}) does not contain damping, which is assumed to be negligibly small, as is the case in many experimental systems.

\section*{Classical spin system}
In order to best illustrate the mechanism of $\PT$ symmetry breaking, we focus on the classical limit, ${S \to \infty}$ and ${k_z\,S \to D/2}$, where $D$ is the dimensionless uniaxial anisotropy constant. Formula~(\ref{zHamilton}) then yields the following equation of motion for the Hamiltonian~(\ref{HDz}):
\begin{equation}\label{solDz}
	\dot z(t) = -\frac{i\,(h_x + \beta)}2\!\left( z^2 - \frac{h_x - \beta}{h_x + \beta}\right) - i\,D\, z\frac{1 \!-\! |z|^2}{1 + |z|^2}\,,
\end{equation}
with up to six fixed points $z_k$, $k = 1, \dots, 6$.

\begin{figure*}[!t]
	\includegraphics[width=2\columnwidth]{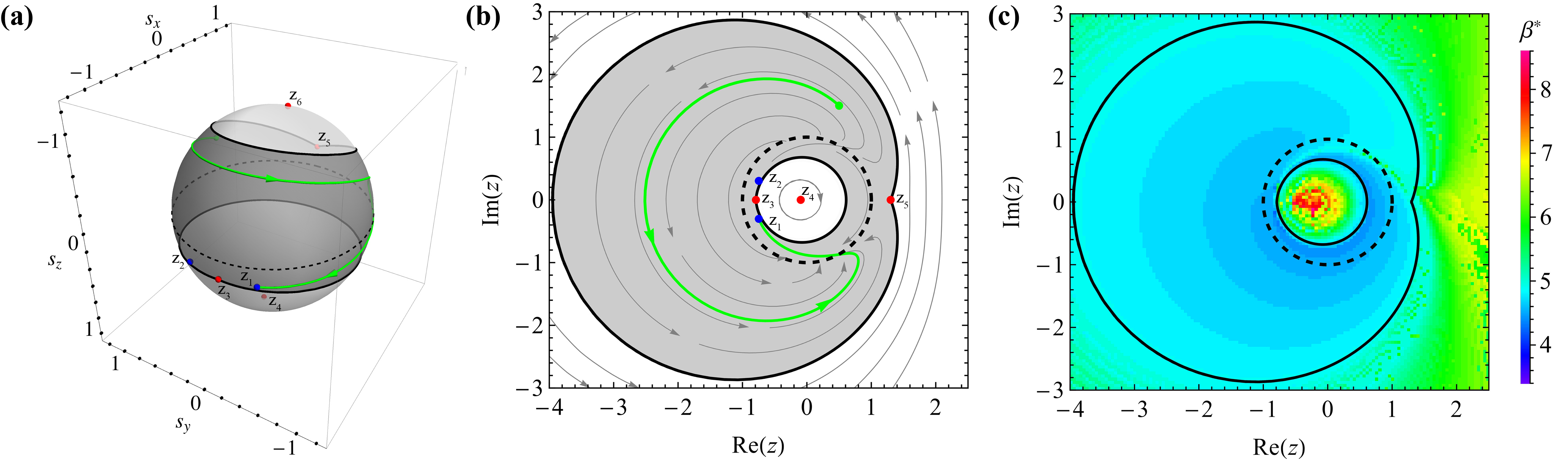}
	\caption{(a, b) Spin dynamics described by Eq.~(\ref{solDz}) with ${h_x = 1}$, ${\beta = 4.7}$, and ${D = 20}$. $\PT$ symmetry is broken in the shaded region around the easy plane ${|z| = 1}$ (dashed line), encompassing two fixed points, $z_{1,2}$ (blue dots), appearing as source and sink nodes. The green line depicts a typical nonoscillatory spin trajectory in the region of broken $\PT$ symmetry. Red dots represent the fixed points $z_{3-6}$. (c) Results of micromagnetic simulations for $\beta^*$ as a function of stereographic projection of the initial spin direction $z$. In the blue region, ${4.6 \lesssim \beta^* \lesssim 4.8}$, the $\PT$ symmetry is broken at all ${|\beta| < \beta^*}$, and the spin takes under $0.5$ ns to saturate in the direction of $z_1$, which is in full agreement with the analytical result.}
	\label{fig3}
\end{figure*}

Shown in Fig.~\ref{fig2} are the real and imaginary parts of the energy spectrum $E_{1-6}$ as functions of the STT amplitude $\beta$. It reveals that in a system with strong anisotropy, ${D \gg 1}$, $\PT$ symmetry breaking occurs in three separate transitions, with the first one at $|\beta| = \beta_1 = |h_x|\sqrt{\bigl[1 + \sqrt{1 + (2D/|h_x|)^2}\,\bigr]/2}$, which corresponds to the smallest amplitude of STT at which ${\Im(E) \ne 0}$. Therefore, $\PT$ symmetry is not broken in the entire phase space of initial spin directions simultaneously, at variance to the linear spin system with ${D = 0}$ (see Appendix B). Instead, the regions of broken and unbroken $\PT$ symmetry may coexist in the phase diagram of a nonlinear spin system.

In what follows we consider a system described by the Hamiltonian~(\ref{HDz}) with ${h_x = 1}$ and ${D = 20}$. For all ${|\beta| < \beta_1 \approx 4.5}$, $\PT$ symmetry is unbroken and the character of spin (magnetization) dynamics is oscillatory in the entire phase diagram, i.e., for all possible initial conditions $z$. At $|\beta| = \beta_1$ the phase transition (first of the three, see Fig.~\ref{fig2}) occurs sharply in a wide region around the easy plane, ${|z| = 1}$, i.e. near the equator of the unit $\S$-sphere, shown in gray in Figs.~\ref{fig3}(a) and \ref{fig3}(b) in Cartesian and stereographic projection coordinates, respectively. It this region the nature of spin dynamics becomes fundamentally different---all spin trajectories follow the lines connecting the fixed points $z_1$ and $z_2$, where ${z_{1, 2} = -\bigl( Dh_x  \pm i \sqrt{\beta^4 - \beta^2 h_x^2 - D^2h_x^2} \bigr)\!/(h_x + \beta)\beta}$, and no closed trajectories are possible; see Fig.~\ref{fig3}(b).

As $|\beta|$ is increased further, the region of broken $\PT$ symmetry expands until it eventually closes around the fixed point $z_5$ at $\beta_2 \approx 9.3$ (second bifurcation in Fig.~\ref{fig2}) and, eventually, the last region of unbroken $\PT$ symmetry near $z_3$ disappears at $\beta_3 \approx 10.8$. The second and third phase transitions are less relevant experimentally as they occur in the vicinity of the least favorable spin directions (parallel and antiparallel to the hard axis $z$) and at considerably higher applied currents.

The predicted transition from precessional dynamics (unbroken $\PT$ symmetry) to exponentially fast saturation in the direction $z_1(h_x, \beta)$ for any initial spin position around the easy plane (broken $\PT$ symmetry) occurs in the setup with mutually perpendicular applied magnetic field and Slonczewski STT. Such a transition in nanoscale magnetic structures can be used for STT- or magnetic-field-controlled magnetization switching in spin valves and a variety of other experimental systems. This effect can further be used for direct measurements of the amplitude of the applied STT, which, unlike the applied current, can be hard to quantify experimentally. 

\section*{Numerical simulations of $\PT$ symmetry breaking}
Here we present the results of numerical simulations confirming the $\PT$ symmetry-breaking phase transition in the classical single-spin system~(\ref{HDz}) by modeling magnetization dynamics of a ferromagnetic disk $100$ nm in diameter and $d = 5$ nm thick, which is consistent with the anisotropy constant $D = 20$ in Eq.~(\ref{solDz}). We used the following typical permalloy material parameters: damping constant ${\alpha = 0.01}$, exchange constant ${A_\text{ex} = 13 \times 10^{-12}}$ J/m and saturation magnetization $M_\text{sat} = 800 \times 10^3$ A/m . The simulations were carried out using the open-source GPU-accelerated micromagnetic simulation program MuMax3~[\onlinecite{mumax3}] based on the LLGS equation~(\ref{LLGSs}) discretized in space. We used a cubic discretization cell of $5$ nm in size, which is smaller than the exchange length in permalloy, $l_\text{ex} = (2A_\text{ex}/\mu_0M_\text{sat}^2)^{1/2} \approx 5.7$ nm.

The permalloy disk was simulated in an external magnetic field applied along the $x$ axis, ${H_0 = 400}$ Oe, which corresponds to the characteristic time ${\tau \approx 0.14}$ ns. The STT was produced by applying electric current perpendicular to the disk in the $z$ direction with spin polarization $\eta = 0.7$ along $\ep = \hat y$ (see Fig.~\ref{fig1}) and current density $\beta$ measured in dimensionless units of $2eH_0M_\text{sat}d/\eta \hbar \approx 0.7 \times 10^8$ A/cm$^2$. While such current density is comparable to typical switching current densities in STT-RAM devices~[\onlinecite{Wang},\onlinecite{Kent04}], its magnitude can be optimized for various practical applications by changing $H_0$ and adjusting the size, shape, and material of the ferromagnetic element.

For all possible initial spin directions $z$, we calculated the critical amplitude of the applied STT, $\beta^*$, for which the character of spin dynamics changes from oscillatory (at $|\beta| < \beta^*$) to exponential saturation. Shown in Fig.~\ref{fig3}(c) is the color map of $\beta^*$ as a function of $z$ in complex stereographic coordinates. The region shown in the shades of blue corresponds to the initial conditions $z$, for which the minimum values of $\beta$ that would guarantee saturation of spin dynamics in the direction of $z_1$ in under $0.5$ ns are between 4.6 and 4.8. This is in full agreement with the region of broken $\PT$ symmetry at ${\beta = 4.7}$ calculated analytically, i.e., the shaded area in Fig.~\ref{fig3}(b) [the outline is repeated in Fig.~\ref{fig3}(c) for comparison]. Outside of this region, a considerably larger magnitude of the applied STT is required to break $\PT$ symmetry.

The agreement between theoretical results and micromagnetic simulations is remarkable considering the non-zero Gilbert damping parameter (${\alpha = 0.01}$) and nonlinear effects (demagnetizing field, finite size and boundary effects, etc.) inherently present in the micromagnetic simulations but not included in the model Hamiltonian~(\ref{HDz}).

\begin{figure*}[!bth]
	\begin{subfigure}
		\centering
		\includegraphics[width=\columnwidth]{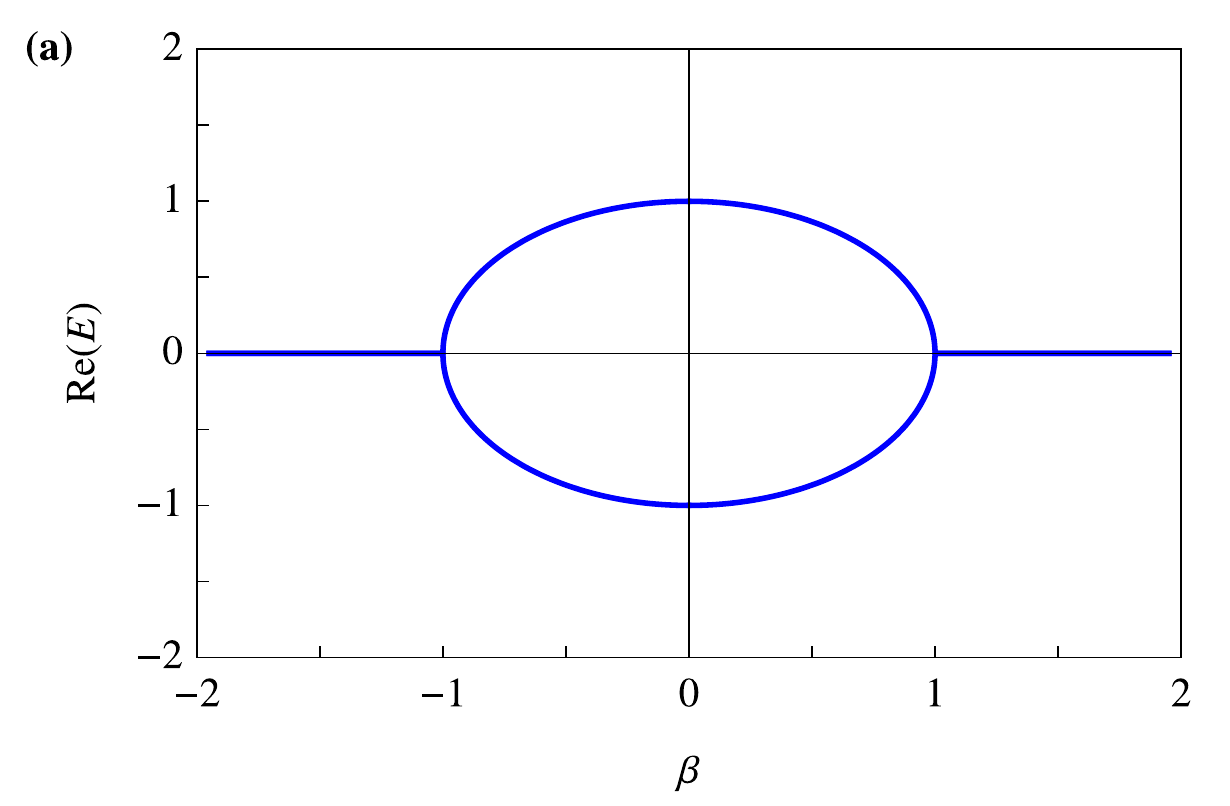}
	\end{subfigure}
	\begin{subfigure}
		\centering
		\includegraphics[width=\columnwidth]{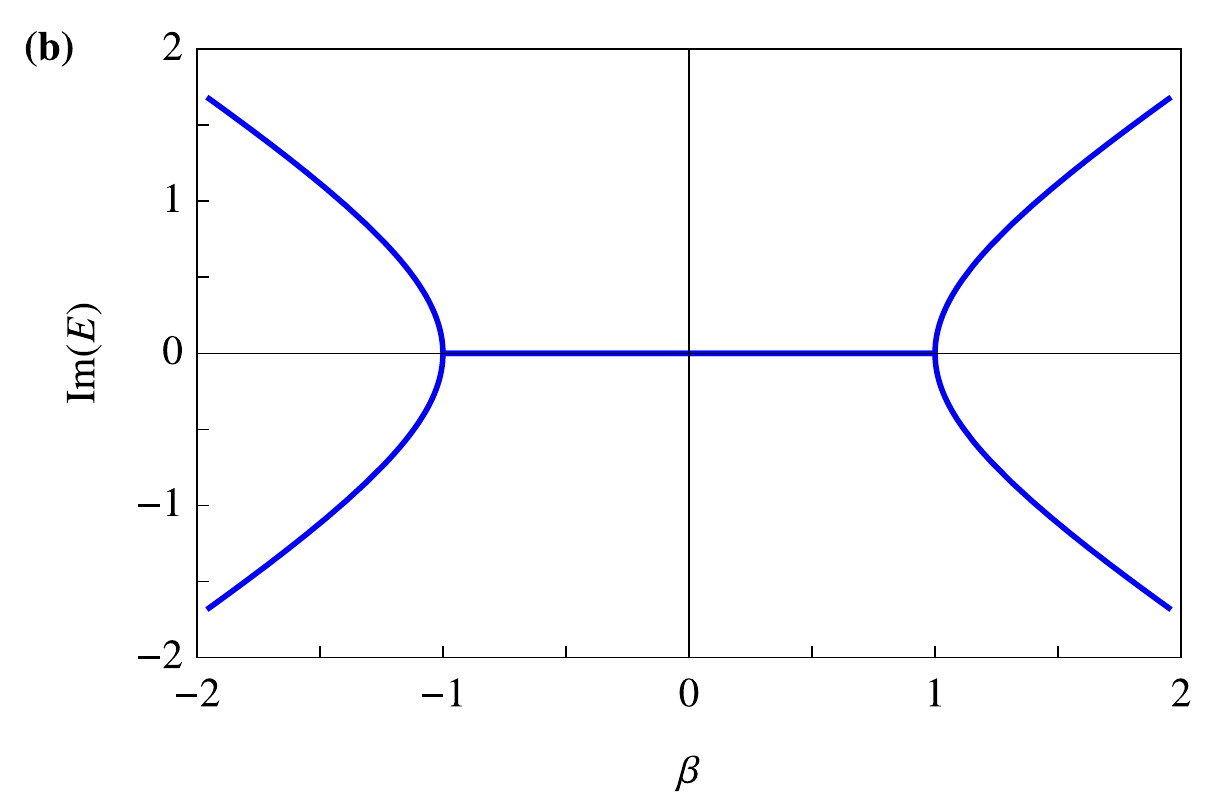}
	\end{subfigure}	
	\caption{Real (a) and imaginary (b) parts of energy spectrum of the linear Hamiltonian $\hat \H_{0\PT}$ as functions of $\beta$ for ${h_x = 1}$. $\PT$ symmetry-breaking transition occurs at ${|\beta| = 1}$.}
	\label{figS1}
\end{figure*}

\section*{Conclusion}
The presented non-Hermitian Hamiltonian formulation of dissipative nonequilibrium spin dynamics generalizes the previous result~[\onlinecite{Wieser}], where the classical Landau-Lifshitz equation was derived from a non-Hermitian Hamilton operator, to open STT-driven spin systems. The introduction of Slonczewski STT in the imaginary part of the Hamiltonian revealed the possibility of STT-driven $\PT$ symmetry-breaking phase transition. Micromagnetic simulations confirm the $\PT$ symmetry-breaking phenomenon in realistic mesoscopic magnetic systems and its robustness against weak dissipation, indicating high potential for impacting spin-based information technology. The way STT enters the complex Hamiltonian~(\ref{H}), i.e. as \textit{imaginary} magnetic field, provides a unique tool for studying Lee-Yang zeros~[\onlinecite{Lee-Yang}] in ferromagnetic Ising and Heisenberg models and, more generally, dynamics and thermodynamics in the complex plane of physical parameters. We envision further realizations of the $\PT$ symmetry-breaking phase transitions in diverse many-body condensed-matter systems and the expansion of practical implementations of the $\PT$ symmetry beyond the present realm of optics~[\onlinecite{Peng}] and acoustics~[\onlinecite{Fleury}].

\section*{Acknowledgements}

We thank Alex Kamenev for critical reading of the manuscript and valuable suggestions. This work was supported by the U.S. Department of Energy, Office of Science, Basic Energy Sciences, Materials Sciences and Engineering Division.

\section*{Appendix A. Generalized equation of motion for non-Hermitian spin Hamiltonians in the classical limit}

The remarkable simplicity of the equation of motion (\ref{zHamilton}) for an arbitrary non-Hermitian spin Hamiltonian function $\H$ stems from the choice of parametrization of the phase space, i.e., the complex stereographic projection coordinates $\{ z, \bar z\}$. The extension of classical equations of motion to non-Hermitian Hamiltonians in terms of canonical coordinates $\{q, p\}$ has the following generalized form~[\onlinecite{Graefe10}]:
\begin{equation}\tag{A1}
	\begin{pmatrix} q \\ p \end{pmatrix} = \Omega^{-1}\nabla_{q, p}(\Re\, \H) - G^{-1}\nabla_{q, p}(\Im\ \H)\,,
\end{equation}
where $\Omega$ and $G$ are the symplectic structure and metric of the underlying classical phase space, respectively, which must satisfy the compatibility condition~[\onlinecite{Brody01}] written in the matrix form as
\begin{equation}\tag{A2}
\Omega^{-1} = \left( \Gamma^{-1}\,\Omega\,\Gamma^{-1}\right)^\text{T}\,.
\end{equation}

In the stereographic projection coordinates, one obtains the following symplectic structure and metric:
\begin{equation}\tag{A3}
	\Omega = \frac{2}{\left( 1 + |z|^2\right)^2} \begin{pmatrix} 0 & i \\ -i & 0\end{pmatrix}\,, \qquad G = \frac{2}{\left( 1 + |z|^2\right)^2} \begin{pmatrix} 0 & 1 \\ 1 & 0\end{pmatrix}\,.
\end{equation}
It is the form of these matrices that leads to Eq.~(\ref{zHamilton}), where the real and imaginary parts (as written in the $\S$ representation) of the Hamiltonian combine naturally into a single complex function $\H$. Therefore, when written in stereographic projection coordinates, the generalized classical equation of motion for non-Hermitian Hamiltonians coincides with that for traditional Hermitian Hamiltonians.

\section*{Appendix B. $\PT$ symmetry breaking in linear spin system}

\begin{figure*}[!tbh]
	\includegraphics[width=1.68\columnwidth]{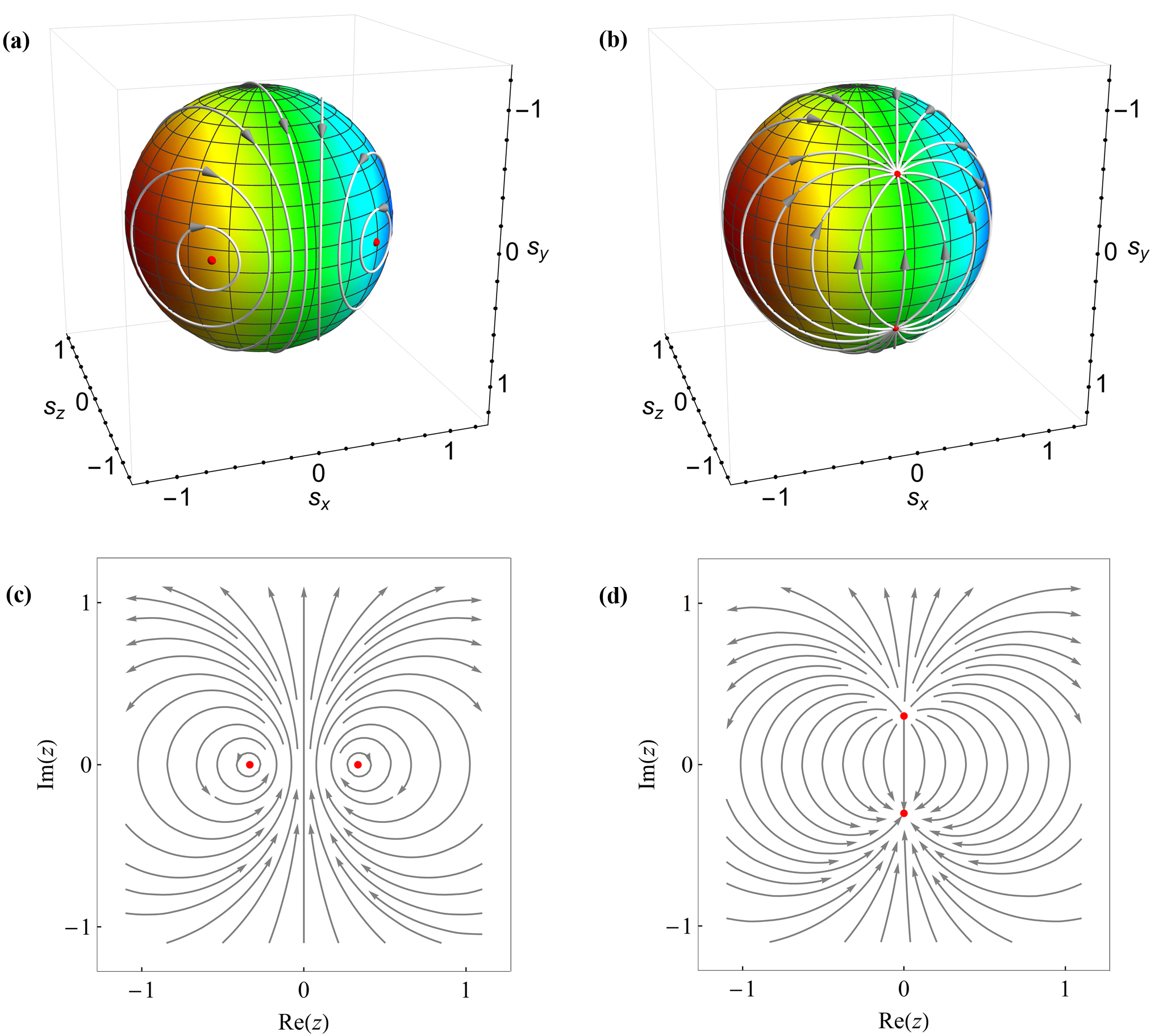}
	\caption{Spin trajectories for classical linear Hamiltonian $\H_{0\PT}$ in the regime of unbroken, (a) and (c), (at $\beta = 0.8$) and broken, (b) and (d), (at $\beta = 1.2$) $\PT$ symmetry for $h_x = 1$.}
	\label{figS2}
\end{figure*}

In the absence of magnetic anisotropy fields, the Hamiltonian~(1) from the main text becomes linear:
\begin{equation}\label{H0}\tag{B1}
	\hat\H_0 = \left( \frac{\gamma\HH + i\,\j}{1 - i\,\alpha}\right) \cdot \hat\S\,,
\end{equation}
with effects of applied magnetic field, damping and Slonczewski STT contributions all incorporated in the \textit{complex} magnetic field (in parentheses).
The $\PT$-symmetric version of this Hamiltonian has mutually perpendicular real and imaginary parts of the complex magnetic field:
\begin{equation}\label{HPT}\tag{B2}
	\hat\H_{0\PT} = \gamma H_0 \left(h_x \hat S_x + i\,\beta\, \hat S_y\right)\,.
\end{equation}

The quantum spin-$\frac12$ version of this Hamiltonian describes a two-level quantum system with balanced loss and gain and is known~[\onlinecite{Bender03},\onlinecite{Brody14}] to exhibit $\PT$ symmetry-breaking transition at ${h_x = \pm\beta}$. When ${|h_x| > |\beta|}$, the Hamiltonian $\hat\H_{0\PT}$ has real eigenvalues, ${\lambda_{1,2} = \pm\sqrt{h_x^2 - \beta^2}}$, while in the parametric region ${|h_x| < |\beta|}$ eigenvalues are imaginary, see Fig.~\ref{figS1}.

The generalized Hamilton's equation of motion~(\ref{zHamilton}) for this Hamiltonian takes the form
\begin{equation}\label{sol}\tag{B3}
\dot z(t) = f(t) = -\frac{i\,(h_x + \beta)}2\left( z^2 - \frac{h_x - \beta}{h_x + \beta}\right)\,.
\end{equation}
It has two fixed points: $z_{1,2} = \pm\sqrt{(h_x - \beta)/(h_x + \beta)}$, which correspond to stereographic coordinates of the spin equilibria directions on a unit Bloch sphere. The character of spin dynamics around the fixed points is fully determined by the eigenvalues of the complex Jacobian matrix ${J_\C = \partial\!\left(\Re\, f, \Im\, f\right)/\partial\!\left(\Re\, z, \Im\, z\right)}$ in their vicinity. The solution of Eq.~(\ref{sol}) takes the form of a M\"{o}bius transformation:
\begin{equation}\tag{B4}
z(t) = \frac{z(0) + i\frac{h_x + \beta}{\sqrt{h_x^2 - \beta^2}}\tan\!\left( \frac{\sqrt{h_x^2 - \beta^2}}2t\right)}{i\frac{h_x - \beta}{\sqrt{h_x^2 - \beta^2}}\tan\!\left( \frac{\sqrt{h_x^2 - \beta^2}}2t\right) z(0) + 1}\,.
\end{equation}

The parametric region ${|h_x| > |\beta|}$ defines the regime of \textit{unbroken} $\PT$ symmetry with real Hamiltonian eigenvalues, ${E_{1,2} = \pm \sqrt{h_x^2 - \beta^2}}$. In the classical approximation, the spin performs persistent oscillations along circular orbits about the fixed points $z_{1,2}$ situated on the real axis, see Figs.~\ref{figS2}(a) and \ref{figS2}(c). The eigenvalues of $J_\C$ at $z_{1,2}$ are purely imaginary, identifying the fixed points are \textit{centers}, according to the standard classification~[\onlinecite{Anosov}]. Closed trajectories represent $\PT$-symmetric dynamics with balanced loss and gain: the spin system gains and loses equal amounts of energy from the nonconservative term $i\,\beta\, S_y$ on the $y < 0$ and $y > 0$ segments of trajectories, respectively.

As the driving parameter $|\beta|$ is increased, $z_{1,2}$ move towards each other until they eventually collide at $|\beta| = |h_x|$, which marks the point of $\PT$ symmetry breaking. In the regime of \textit{broken} $\PT$ symmetry, ${|h_x| < |\beta|}$, the energy eigenvalues are imaginary, ${E_{1,2} = \pm i\sqrt{\beta^2 - h_x^2}}$, and no closed trajectories are possible. The eigenvalues of the Jacobian $J_\C(z_{1,2})$ in this regime are real and of the same sign, defining $z_{1,2}$ as \textit{sink} and \textit{source nodes}~[\onlinecite{Anosov}]. All trajectories follow circle arcs connecting the fixed points with coordinates in three-dimensional space ${\left(0, \pm\sqrt{1 - (h_x/\beta)^2}, -h_x/\beta\right)}$, which are now out of the $xz$ plane, see Figs.~\ref{figS2}(b) and \ref{figS2}(d).

We emphasize that in the linear Hamiltonian~(\ref{H0}) the effects of damping can always be fully compensated by the appropriate choice of the applied STT. For instance, the $\PT$-symmetric Hamiltonian~(\ref{HPT}) describes a single spin placed in external magnetic field ${\HH = H_0\left(h_x, \alpha\, \beta, 0\right)}$ and STT ${\j = \gamma H_0\left(-\alpha\, h_x, \beta, 0\right)}$. Note that in the general case of nonlinear spin Hamiltonian, dissipation cannot be completely canceled by STT. However, in many magnetic systems dissipative forces are extremely weak, which justifies the approximation of zero damping.

\section*{Appendix C. Integrals of motion}
There exists a remarkable similarity between spin dynamics in the regime of unbroken $\PT$ symmetry and that of a fully Hermitian system. Indeed, it follows from Eq.~(\ref{sol}) that all spin trajectories are circular with the precession frequency equal to $\sqrt{h_x^2 - \beta^2}$. The equivalent~[\onlinecite{Ali}] to~(\ref{HPT}) Hermitian Hamiltonian reads:
\begin{equation}\label{HHerm}\tag{B5}
\H' = \gamma H_0\sqrt{h_x^2 - \beta^2}\, S_x'\,,
\end{equation}
for which the equation of motion is obtained from Eq.~(\ref{sol}) by the circle-preserving M\"{o}bius transformation 
\begin{equation}\label{transform}\tag{B6}
z' = \sqrt{\frac{h_x + \beta}{h_x - \beta}}\, z\,.
\end{equation}
The relation between time evolution of complex linear spin Hamiltonians and M\"{o}bius transformations will be described elsewhere. The existence of this transformation leads to the non-Hermitian system~(\ref{HPT}) having an integral of motion:
\begin{equation}\tag{B7}
I[z, \bar z] = (h_x + \beta) \frac{z + \bar z}{1 + \frac{h_x + \beta}{h_x - \beta}|z|^2}\,,
\end{equation}
which is nothing but the magnetic energy conserved by the Hermitian Hamiltonian~(\ref{HHerm}):
\begin{equation}\tag{B8}
I\!\left[ z', \bar z'\right] = \sqrt{h_x^2 - \beta^2}\, s'_x = \sqrt{h_x^2 - \beta^2}\frac{z' + \bar z'}{1 + |z'|^2}
\end{equation}
after the transformation (\ref{transform}).

\bibliographystyle{apsrev4-1}

\end{document}